\documentclass[aps,pra,twocolumn,showpacs,groupedaddress,preprintnumbers,amsmath,amssymb]{revtex4-1}
\usepackage{mathrsfs}

\usepackage{graphicx}
\usepackage{bm}
\usepackage{float}

\begin{document}

\preprint{APS/123-QED}

\title{Quantum phase transition of light in a 1-D photon-hopping-controllable resonator~array}

\author{Chun-Wang Wu}
  \email{cwwu@nudt.edu.cn}
\author{Ming Gao}
\author{Zhi-Jiao Deng}
\author{Hong-Yi Dai}
\author{Ping-Xing Chen}
\author{Cheng-Zu Li}
\affiliation{College of Science, National University of Defense
 Technology, Changsha 410073, People's Republic of China}

\date{\today}

\begin{abstract}
 We give a concrete experimental scheme for engineering the insulator-superfluid transition of light
 in a one-dimensional (1-D) array of coupled superconducting stripline resonators. In our proposed
 architecture, the on-site interaction and the photon hopping rate can be tuned independently by adjusting
 the transition frequencies of the charge qubits inside the resonators and at the resonator junctions, respectively,
 which permits us to systematically study the quantum phase transition of light in a complete parameter space.
 By combining the techniques of photon-number-dependent qubit transition
 and fast read-out of the qubit state using a separate low-Q resonator mode, the statistical property of
 the excitations in each resonator can be obtained with a high efficiency. An analysis of the
 various decoherence sources and disorders shows that our scheme can serve as a guide to coming experiments
 involving a small number of coupled resonators.
\end{abstract}


\maketitle

\section{introduction}

In the past two decades, there has been a great interest in mimicking various quantum many-body phenomena
with artificially engineered structures that permit unprecedented experimental control and measurement
access \cite{reve1,reve2}. A very successful example in this direction is the simulation of Mott
insulator-superfluid transition with ultracold atoms in optical lattices \cite{reve3}. Recently, the
coupled resonator array has been suggested as another promising candidate for building a quantum simulator
\cite{reve4,reve5}. Compared to other structures, the coupled resonator array can be used to study the
quantum phase transition of light and has the striking advantage of full addressability of individual sites.

Since the idea of realizing strongly correlated states of light in coupled resonator arrays was first proposed
in the seminal papers \cite{reve6,reve7,reve8}, a large amount of work has been devoted to a systematic study
of the light phase in the total parameter space \cite{reve9,reve10,reve11,reve12,reve13,reve14,reve15,reve16,reve17,reve18,reve19}.
The phase boundary between the Mott insulator (MI) phase and the superfluid (SF) phase has been obtained using
different numerical methods including the mean-field approach \cite{reve9,reve10}, the quantum Monte Carlo
simulation \cite{reve11}, and the density-matrix renormalization-group approach \cite{reve12,reve13}. Analytical
and numerical methods were also developed to calculate the single-particle excitation spectrum of the phase
space \cite{reve14,reve15,reve16}. By considering effects of the dissipation and driving terms, recent work
by several groups has promoted the quantum phase transition of light to a nonequilibrium case \cite{reve17,reve18,reve19}.

Despite the plentiful and substantial achievements in the theoretical aspect, there have not yet been any experimental
realizations of coupled resonator arrays. For an experimental exploration of the quantum phase transition of light
using coupled resonator arrays, we should have the abilities of preparing the total system in its ground state, tuning
the effective photon repulsion and the photon hopping rate over a wide range of values, and obtaining the accurate
statistical property of the excitations in each resonator. Up to now, previous work has just provided some primitive
hints toward possible realizations of the model \cite{reve4,reve5,reve20}. The authors in Refs. \cite{reve7,reve8} suggested
measuring the individual resonator via mapping the excitations onto the atomic levels followed by state selective resonance
fluorescence, but this method still suffers from the current lack of high-efficiency photon detectors. Moreover, the photon
hopping rate is typically fixed by the fabrication process and can not be changed post-creation, which limits the study of
quantum phase transition of light in a complete parameter space.

In this paper, we propose an alternative experimental scheme which does not suffer from the above limitations. For our
version of the coupled resonator array, each superconducting transmission line resonator (TLR) contains a charge qubit as the
nonlinear element and the adjacent resonators are coupled by another charge qubit playing the role of an effective knob for the
photon hopping rate. The local statistical property of each superconducting resonator can be analyzed readily using combined
techniques of photon-number-dependent qubit transition \cite{reve21, reve22} and fast read-out of the qubit state through a
separate low-Q resonator mode \cite{reve23}, for which the high-efficiency photon detectors are not required. Because all
the techniques we use have been separately demonstrated in the laboratory, our scheme may be implemented in the near future.

\begin{figure*}[!t]
 \includegraphics[scale=0.5]{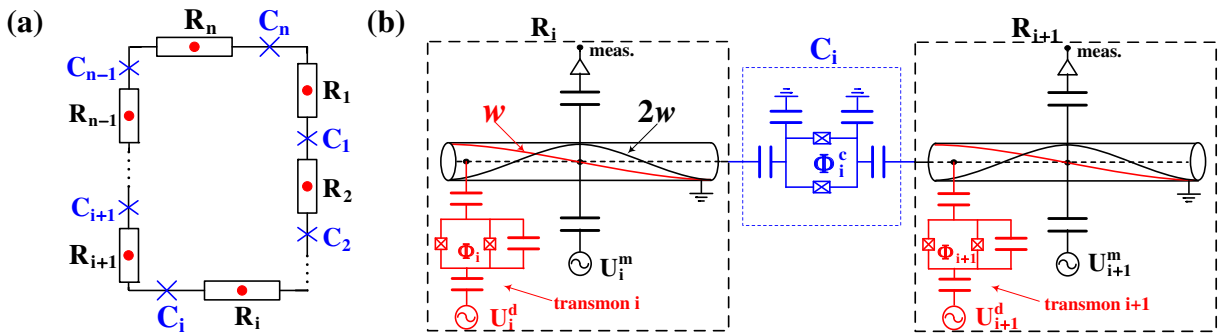}
 \setlength{\belowcaptionskip}{-0.4cm}
 \caption{\label{fig1}(Color online) (a) Schematic layout of our proposed architecture. A 1-D array of $n$ TLRs $R_{1},R_{2},\,...\,,R_{n}$
 is arranged into an annular geometry to satisfy the periodic boundary conditions. Each resonator contains a charge qubit to offer the strong nonlinearity.
 The adjacent resonators $R_{i}$ and $R_{i+1}$ are coupled by another charge qubit $C_{i}$ playing the role of an effective knob for the photon hopping rate.
 (b) Electrical circuit representation of the adjacent resonators and their junction. The high-Q half-wave mode of $R_{i}$ with resonance frequency $w$ is
 coupled to transmon $i$, and $R_{i}$'s low-Q full-wave mode with resonance frequency $2w$ is strongly coupled to a measurement line fabricated at
 the resonator center. The state of transmon $i$ can be measured by applying a microwave field $U^{m}_{i}$ at the input port of the measurement line. Another
 transmon $C_{i}$ is injected into the resonator junction , which is dispersively coupled to $R_{i}$ and $R_{i+1}$. Microwave pulse $U^{d}_{i}$ applied to
 the gate serves for driving the qubit transition of transmon $i$. The transition frequency of each transmon can be tuned via the corresponding applied magnetic
 flux $\Phi$. }
 \end{figure*}

The paper is organized as follows. In Sec.\ \uppercase\expandafter{\romannumeral2} we introduce the physical system considered
and derive the effective Hamiltonian. In Sec.\ \uppercase\expandafter{\romannumeral3}, a static phase diagram in the complete
parameter space is obtained using the exact diagonalization techniques and some related results are discussed. Finally, we
explore the possibilities of engineering photonic MI-SF transition dynamically using our proposed architecture in Sec.\
\uppercase\expandafter{\romannumeral4}. A concrete experimental procedure is presented including how to initialize the system,
tune the system's parameters, and measure the individual resonators. The various decoherence sources and disorders are also analyzed.

 \section{the physical system and effective Hamiltonian}

The system we consider is schematically depicted in Fig.\,1(a). A 1-D array of $n$ TLRs $R_{1},R_{2},\,...\,,R_{n}$
 is arranged into an annular geometry to satisfy the periodic boundary conditions. Each resonator contains a charge qubit (represented
 by the red dot) to offer the strong nonlinearity. The adjacent resonators $R_{i}$ and $R_{i+1}$ are coupled by another charge qubit
 $C_{i}$ (represented by the blue cross) playing the role of an effective knob for the photon hopping rate. The charge qubit used in
 our model is transmon, a modified version of the Cooper pair box proposed by Koch \emph{et al.}\cite{reve24}. Its unique feature
 is the large shunt capacitor between the superconducting islands, which makes the transmon have a longer decoherence time than the
 ordinary charge qubits. High-Q resonators are advantageous for the simulation of MI-SF transition but are adverse to the measurement
 of individual sites. To solve this problem, we can use the technique of engineering two modes of a resonator with different quality
 factors, which has been demonstrated experimentally in \cite{reve23}. As shown in Fig.\,1(b), the high-Q half-wave mode of $R_{i}$
 with resonance frequency $w$ is coupled to transmon $i$, and $R_{i}$'s low-Q full-wave mode with resonance frequency $2w$
 is strongly coupled to a measurement line fabricated at the resonator center. The state of transmon $i$ can be measured by applying
 a microwave field $U^{m}_{i}$ at the input port of the measurement line. Another transmon $C_{i}$ is injected into the resonator
 junction, which is dispersively coupled to $R_{i}$ and $R_{i+1}$. Note that the usual shunt capacitor between the superconducting
 islands of $C_{i}$ is replaced with capacitors to the ground planes to suppress direct coupling between $R_{i}$ and $R_{i+1}$ \cite{reve22}. Microwave
 pulse $U^{d}_{i}$ applied to the gate serves for driving the qubit transition of transmon $i$. The transition frequency of each transmon
 can be tuned via the corresponding applied magnetic flux $\Phi$.

 In the following, we will derive the effective Hamiltonian of our proposed architecture. Let us denote the lowest two eigenstates of transmon
 $i$ with $|g\rangle_{i}$ and $|e\rangle_{i}$, which are separated by energy $\epsilon$ and coupled to the half-wave mode of $R_{i}$ with qubit-resonator
 coupling strength $g$. $C_{i}$'s lowest two eigenstates $|g^{c}\rangle_{i}$ and $|e^{c}\rangle_{i}$, separated by energy $\epsilon_{c}$, are simultaneously
 coupled to $R_{i}$ and $R_{i+1}$ with coupling strength $g_{c}$. In this paper, our analysis is restricted to the case of $\epsilon_{c}-w\gg g_{c}$,
 i.\,e.\,, $C_{i}$ is dispersively coupled to its neighbor resonators. The Hamiltonian for the total system can be written as (assuming $\hbar=1$)
 \begin{equation}
 H=H_{1}+H_{2},\nonumber
 \end{equation}
 \begin{equation}
 H_{1}=\sum_{i=1}^{n}[\epsilon|e\rangle_{ii}\langle e|+wa_{i}^{\dag}a_{i}+g(\sigma_{i}^{+}a_{i}+\sigma_{i}^{-}a_{i}^{\dag})],\nonumber
 \end{equation}
 \begin{equation}
 H_{2}=\sum_{i=1}^{n}[\epsilon_{c}|e^{c}\rangle_{ii}\langle e^{c}|+g_{c}(\sigma_{ci}^{+}a_{i}+\sigma_{ci}^{+}a_{i+1}+H.c.)],
 \end{equation}
 where $H_{1}$ is the sum of local Jaynes-Cummings Hamiltonians with resonator index $i$, photon creation (annihilation) operator $a_{i}^{\dag}$
 ($a_{i}$) and qubit raising (lowering) operator $\sigma_{i}^{+}$ ($\sigma_{i}^{-}$) for transmon $i$; $H_{2}$ describes the sum of interactions
 between $C_{i}(i=1,2,\ldots,n)$ and their neighbor resonators with $\sigma_{ci}^{+}$ ($\sigma_{ci}^{-}$) being $C_{i}$'s qubit raising (lowering)
 operator. Considering the annular geometry of the coupled resonator array, we have $a_{n+1}=a_{1}$ and $a_{n+1}^{\dag}=a_{1}^{\dag}$, which offer the
 periodic boundary conditions for our architecture.

 In an interaction picture with respect to $H_{0}=\sum_{i=1}^{n}[w(|e\rangle_{ii}\langle e|+a_{i}^{\dag}a_{i})+\epsilon_{c}|e^{c}\rangle_{ii}\langle e^{c}|]$, the system Hamiltonian
 reads
 \begin{equation}
 H^{int}=H_{1}^{int}+H_{2}^{int},\nonumber
 \end{equation}
 \begin{equation}
 H_{1}^{int}=\sum_{i=1}^{n}[\Delta|e\rangle_{ii}\langle e|+g(\sigma_{i}^{+}a_{i}+\sigma_{i}^{-}a_{i}^{\dag})],\nonumber
 \end{equation}
 \begin{equation}
 H_{2}^{int}=\sum_{i=1}^{n}[g_{c}\sigma_{ci}^{+}(a_{i}+a_{i+1})e^{i\Delta_{c}t}+g_{c}\sigma_{ci}^{-}(a_{i}^{\dag}+a_{i+1}^{\dag})e^{-i\Delta_{c}t}],
 \end{equation}
 where $\Delta=\epsilon-w$, and $\Delta_{c}=\epsilon_{c}-w$. With the choice of $\Delta_{c}\gg g_{c}$, the real energy exchanges between $C_{i}(i=1,2,\ldots,n)$ and their neighbor resonators are largely suppressed. In this case, we can use the time-averaging method in Ref.\,\cite{reve25} and neglect the effect of rapidly oscillating terms. Then $H_{2}^{int}$ can be approximated by
 \begin{eqnarray}
 H_{2}^{'}&=&\sum_{i,j=1}^{n}\frac{1}{\Delta_{c}}[g_{c}\sigma_{ci}^{+}(a_{i}+a_{i+1}),g_{c}\sigma_{cj}^{-}(a_{j}^{\dag}+a_{j+1}^{\dag})] \nonumber \\
  &=&\sum_{i=1}^{n}\frac{g_{c}^{2}}{\Delta_{c}}[\sigma_{ci}^{z}(a_{i}^{\dag}a_{i}+a_{i+1}^{\dag}a_{i+1})+2|e^{c}\rangle_{ii}\langle e^{c}| \nonumber \\
  &&+\sigma_{ci}^{z}(a_{i+1}^{\dag}a_{i}+H.c.)+(\sigma_{ci}^{+}\sigma_{ci+1}^{-}+H.c.)],
 \end{eqnarray}
 where $\sigma_{ci}^{z}=|e^{c}\rangle_{ii}\langle e^{c}|-|g^{c}\rangle_{ii}\langle g^{c}|$. The four terms of $H_{2}^{'}$ describe the ac Stark shifts, Lamb shifts, transmon-intermediated photon
 hoppings, and photon-intermediated dipole couplings between the transmons situated at the neighbored junctions, respectively. If $C_{i}(i=1,2,\ldots,n)$ are prepared
 in $|g^{c}\rangle_{1}|g^{c}\rangle_{2}\ldots|g^{c}\rangle_{n}$, then they will always stay in their ground states. In this situation, $H_{2}^{'}$ can be simplified by
 \begin{eqnarray}
 H_{2}^{''}&=&_{n}\langle g^{c}|\ldots_{2}\langle g^{c}|_{1}\langle g^{c}|H_{2}^{'}|g^{c}\rangle_{1}|g^{c}\rangle_{2}\ldots|g^{c}\rangle_{n}  \nonumber \\
  &=&\sum_{i=1}^{n}[-\frac{g_{c}^{2}}{\Delta_{c}}(a_{i+1}^{\dag}a_{i}+a_{i}^{\dag}a_{i+1})-\frac{2g_{c}^{2}}{\Delta_{c}}a_{i}^{\dag}a_{i}].
 \end{eqnarray}
 By moving to a second interaction picture with respect to $\sum_{i=1}^{n}-\frac{2g_{c}^{2}}{\Delta_{c}}(a_{i}^{\dag}a_{i}+|e\rangle_{ii}\langle e|)$, the total system
 Hamiltonian $H_{1}^{int}+H_{2}^{''}$ yields
 \begin{equation}
 H^{eff}=H^{hop}+H^{repul},\nonumber\\
 \end{equation}
 \begin{equation}
 H^{hop}=\sum_{i=1}^{n}-\kappa(\Delta_{c})(a_{i+1}^{\dag}a_{i}+a_{i}^{\dag}a_{i+1}),\nonumber\\
 \end{equation}
 \begin{equation}
 H^{repul}=\sum_{i=1}^{n}[\Delta^{'}|e\rangle_{ii}\langle e|+g(\sigma_{i}^{+}a_{i}+\sigma_{i}^{-}a_{i}^{\dag})],
 \end{equation}
 where $\Delta^{'}=\Delta+\frac{2g_{c}^{2}}{\Delta_{c}}$, and $\kappa(\Delta_{c})=\frac{g_{c}^{2}}{\Delta_{c}}$ is the $\Delta_{c}$-dependent photon hopping rate.

 The Hamiltonian $H^{repul}$ in Eq.\,(5) actually provides an effective ($\Delta,\Delta_{c}$)-dependent on-site repulsion for excitations. $H^{repul}$ can be
 diagonalized in a basis of mixed photonic and atomic excitations, called polaritons\cite{reve6,reve8}. Let $|n,g\rangle$ ($|n,e\rangle$) represent a resonator that contains $n$ photons and a single qubit in the ground (excited) state. Then the polariton states of $R_{i}$, labeled by the polariton number $n$ and upper or lower branch index $\sigma=\pm$, can be given by
 \begin{equation}
 |n,+\rangle_{i}=\sin\theta_{n}|n-1,e\rangle_{i}+\cos\theta_{n}|n,g\rangle_{i},\nonumber
 \end{equation}
 \begin{equation}
 |n,-\rangle_{i}=\cos\theta_{n}|n-1,e\rangle_{i}-\sin\theta_{n}|n,g\rangle_{i},
 \end{equation}
 with the mixing angle $\tan\theta_{n}=\frac{\frac{\Delta^{'}}{2}+\sqrt{(\frac{\Delta^{'}}{2})^{2}+ng^{2}}}{\sqrt{n}g}$. The corresponding eigenvalues are
 \begin{equation}
 E_{n}^{\sigma}=\frac{\Delta^{'}}{2}+\sigma\sqrt{(\frac{\Delta^{'}}{2})^{2}+ng^{2}},\qquad \sigma=\pm.
 \end{equation}
 The zero-polariton state $|0,-\rangle_{i}=|0,g\rangle_{i}$ is a special case with $E_{0}^{-}=0$. Obviously, these polariton states are also eigenstates of the
 polariton number operator $N_{i}=a_{i}^{\dag}a_{i}+|e\rangle_{ii}\langle e|$ with eigenvalue $n$. For the case of $\kappa(\Delta_{c})$ not much larger than $g$,
 if we create only the lower polariton states in the resonator array initially, then the upper polariton states will never be created because the interconversion
 between the two polariton branches can be neglected. The effective on-site repulsion $U_{eff}$ for polaritons results from the anharmonicity in the spectrum of $H^{repul}$, which is dependent on the number of polaritons in the resonator. For polariton number $n=1$, the effective repulsion $U_{eff}(1)$ can be identified by calculating the energy cost to inject a second polariton into the resonator,
 \begin{eqnarray}
 U_{eff}(1)&=&E_{2}^{-}-2E_{1}^{-} \nonumber \\
 &=&-\frac{\Delta^{'}}{2}+\sqrt{{\Delta^{'}}^{2}+4g^{2}}-\sqrt{(\frac{\Delta^{'}}{2})^{2}+2g^{2}}.
 \end{eqnarray}
 $U_{eff}(1)$ can be tuned easily by changing the detuning $\Delta^{'}=\Delta+\frac{2g_{c}^{2}}{\Delta_{c}}$. If $\Delta^{'}>0$ and $\Delta^{'}\gg g$, $U_{eff}(1)$
 vanishes; if $\Delta^{'}<0$ and $-\Delta^{'}\gg g$, then $U_{eff}(1)\approx-\Delta^{'}$ is a large quantity.

 \section{static phase diagram in the $(\Delta,\Delta_{c})$ parameter space}

 In Sec.\,\uppercase\expandafter{\romannumeral2}, we have derived the effective Hamiltonian $H^{eff}$ for the proposed architecture,
 which has two in situ tunable parameters $\Delta$ and $\Delta_{c}$. In different regimes of the $(\Delta,\Delta_{c})$ plane, the system
 can show distinct characteristics. For simplicity, our analysis is restricted to the case of the resonator array contains on average
 one polariton per resonator.

 If $\Delta+\frac{2g_{c}^{2}}{\Delta_{c}}=0$ and $\frac{g_{c}^{2}}{\Delta_{c}}\ll g$, the system Hamiltonian is dominated by the resonant
 Jaynes-Cummings interaction, i.e.\,$H^{eff}\approx\sum_{i=1}^{n}g(\sigma_{i}^{+}a_{i}+\sigma_{i}^{-}a_{i}^{\dag})$. If one local resonator
 has a polariton in it, the strong qubit-resonator interaction will shift the frequency of the resonator mode and prevent a second polariton
 from entering it \cite{reve26}. This anharmonicity in the spectrum leads to an effective polariton-polariton repulsion
 $U_{eff}(1)=(2-\sqrt{2})g\gg\kappa(\Delta_{c})=\frac{g_{c}^{2}}{\Delta_{c}}$. In this case, the ground state of the system is approximately
 \begin{equation}
 |\Psi\rangle_{MI}=\otimes_{i=1}^{n}|1,-\rangle_{i},
 \end{equation}
 which can be seen as the MI state of polaritons.

 If $\Delta^{'}=\Delta+\frac{2g_{c}^{2}}{\Delta_{c}}\gg g$, we have $|n,-\rangle_{i}\approx-|n,g\rangle_{i}$ and $U_{eff}(1)\approx0$. When
 $\kappa(\Delta_{c})=\frac{g_{c}^{2}}{\Delta_{c}}$ is nonvanishing, the photon hopping terms will dominate the system Hamiltonian, i.e.\,$H^{eff}\approx H^{hop}=-\sum_{i=1}^{n}\frac{g_{c}^{2}}{\Delta_{c}}(a_{i+1}^{\dag}a_{i}+a_{i}^{\dag}a_{i+1})$. In this situation, $H^{eff}$
 can be diagonalized through the Fourier transform. Introducing $b_{k}=\sum_{j=1}^{n}\frac{1}{\sqrt{n}}e^{i\frac{2\pi jk}{n}}a_{j} \quad (k=0,1,2,\,\ldots\,,n-1)$,
 we have $[b_{k},b_{k}^{\dag}]=1$ and $H^{eff}$ can be rewritten as $H^{eff}\approx H^{hop}=-2\frac{g_{c}^{2}}{\Delta_{c}}\sum_{k=0}^{n-1}\cos\frac{2\pi k}{n}b_{k}^{\dag}b_{k}$. Obviously, the system ground state is given by
 \begin{eqnarray}
 |\Psi\rangle_{SF}&=&\frac{1}{\sqrt{n!}}(b_{0}^{\dag})^{n}|vac\rangle \nonumber \\
 &=&\frac{1}{\sqrt{n!}}(\frac{1}{\sqrt{n}}\sum_{i=1}^{n}a_{i}^{\dag})^{n}|vac\rangle
 \end{eqnarray}
 and the corresponding ground state energy is $-2n\frac{g_{c}^{2}}{\Delta_{c}}$, where $|vac\rangle$ is the vacuum state of the resonator array.
 With $n$ photons delocalizing throughout the whole resonator array, $|\Psi\rangle_{SF}$ is the SF phase state of light.

 By calculating the system ground states corresponding to different values of $\Delta$ and $\Delta_{c}$, we can obtain a static phase diagram
 in the ($\Delta,\Delta_{c}$) plane and observe the phase boundary between the MI phase and the SF phase. In this section, we will show that, even with
 a very small resonator array, the phase diagram can exhibit the main features of MI-SF transition, which is very advantageous
 for the potential experimental realization. To plot the phase diagram, we must choose an order parameter to differentiate between insulatorlike and
 superfluidlike states. For small size resonator arrays, a good order parameter is the variance of the polariton number in a given site \cite{reve7,reve8},
 which is defined as
 \begin{equation}
 var(N_{i})=<N_{i}^{2}>-<N_{i}>^{2}.
 \end{equation}
 In the MI phase, the number of polaritons per resonator is well defined and has zero variance. However, in the SF phase, the polariton number fluctuates and
 thus the variance has a nonzero value. Obviously, for $|\Psi\rangle_{MI}$ in Eq.\,(9), $var(N_{i})=0$; in the case of $n=3$, the SF state $|\Psi\rangle_{SF}$
 in Eq.\,(10) results in a nonvanishing variance $var(N_{i})\approx0.6668$.

 \begin{figure}[!b]
 \includegraphics[scale=0.63]{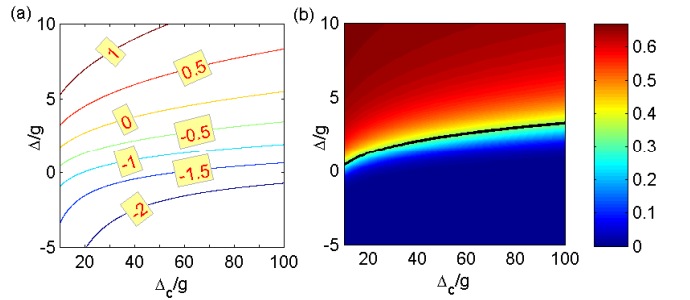}
 \setlength{\belowcaptionskip}{-0.4cm}
 \caption{\label{fig2}(Color online) (a) Contour lines of $\log_{10}(\kappa/U_{eff}(1))$ in the ($\Delta,\Delta_{c}$) parameter space. The system parameters
 we choose are $g=g_{c}$, $\Delta\in[-5g,10g]$, and $\Delta_{c}\in[10g,100g]$. (b) Order parameter $var(N_{i})$ of the system ground state as a function
 of $\Delta$ and $\Delta_{c}$ for a 3-site resonator array. The phase boundary between MI and SF can be approximated by the contour line $\kappa/U_{eff}(1)=0.28$(black solid line). }
 \end{figure}

 The MI-SF phase transition results from the interplay between the photon hopping and the on-site repulsive interaction. As shown in Fig.\,2(a), if we choose the
 system parameters $g=g_{c}$, $\Delta\in[-5g,10g]$ and $\Delta_{c}\in[10g,100g]$, the ratio of the photon hopping rate to the on-site repulsion $\kappa/U_{eff}(1)$
 can be tuned from smaller than $10^{-2}$ to larger than $10$, which permits us to systematically study the phase transition in the ($\Delta,\Delta_{c}$) plane. Note
 that our choice of $\Delta_{c}$ satisfies the dispersive coupling condition $\Delta_{c}\gg g_{c}$, which has been used in deriving the effective Hamiltonian Eq.\,(5).

 For a 3-site resonator array containing on average one polariton per resonator, the system basis consists of state vectors of the form $|\phi\rangle_{k}=\otimes_{i=1}^{3}|n_{i}^{k},s_{i}^{k}\rangle_{i}$, where $n_{i}^{k}\in\{0,1,2,3\}$, $s_{i}^{k}\in\{g,e\}$ and $_{k}\langle\phi|(\sum_{i=1}^{3}N_{i})|\phi\rangle_{k}=3$. It is easy to check that there are a total of $38$ state vectors in this form. To plot the phase diagram in
 ($\Delta,\Delta_{c}$) plane, we must calculate the order parameter $var(N_{i})$ of the system ground state as a function of $\Delta$ and $\Delta_{c}$ as follows.
 First: obtain the matrix elements of $H^{eff}$ in the system basis vectors $|\phi\rangle_{k}\;(k=1,2,\ldots,38)$. Second: diagonalize this $38\times38$ matrix and
 identify the eigenvector $|\phi\rangle_{g}$ corresponding to the lowest eigenvalue. Third: compute the order parameter
 $var(N_{i})=_{g}\langle\phi|N_{i}^{2}|\phi\rangle_{g}-(_{g}\langle\phi|N_{i}|\phi\rangle_{g})^{2}$. The obtained values of $var(N_{i})$ as a function of $\Delta$ and $\Delta_{c}$ are plotted in Fig.\,2(b). It is shown that, although the size of our resonator array is very small, the phase transition takes place over a narrow variation of the parameters. Similar to the Bose-Hubbard model, the value of $var(N_{i})$ is mainly determined by the ratio of the photon hopping rate to the on-site repulsion \cite{reve27}. For our 3-site resonator array, the phase boundary between MI and SF can be approximated by the contour line $\kappa/U_{eff}(1)=0.28$. By fixing $\Delta$ at a certain value and varying $\Delta_{c}$, or fixing $\Delta_{c}$ and varying $\Delta$, we are all able to tune the system from MI phase to SF phase. However, the latter method is more effective because $var(N_{i})$ is more sensitive to $\Delta$.
  
 The phase diagram presented above is obtained using the effective Hamiltonian $H^{eff}$, which neglects the real energy exchanges between $C_{i}$ and their neighbor resonators. Because the off-resonant transitions of $C_{i}$ occur with the probability $\sim$ $(\frac{g_{c}}{\Delta_{c}})^{2}$, the Hamiltonian $H^{eff}$ becomes accurate when $\Delta_{c}$ increases. In Fig.\,3, we compare the order parameters of the approximate and exact system ground states for a 3-site resonator array, obtained from numerical diagonalization of the effective Hamiltonian $H^{eff}$ and the full Hamiltonian $H$ in Eq.\,(1), respectively. The parameters we choose are $g=g_{c}$, $\Delta_{c}=10g_{c}$, and $\Delta\in[-5g, 10g]$. The good agreement between these results indicates that $\Delta_{c}=10g_{c}$ is large enough to guarantee the accuracy of $H^{eff}$.
 
 \begin{figure}[!h]
 \includegraphics[scale=0.63]{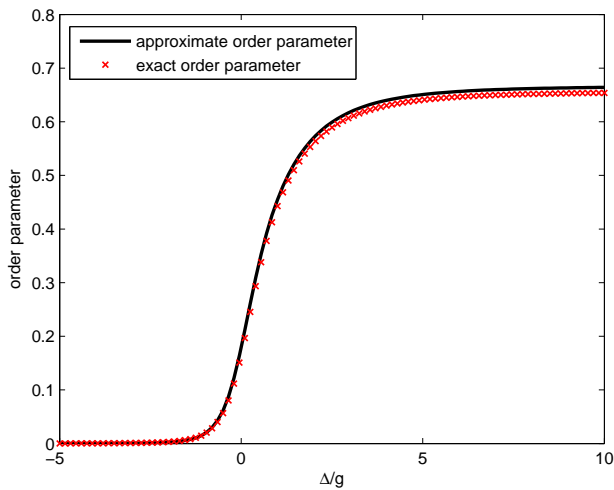}
 \setlength{\belowcaptionskip}{-0.4cm}
 \caption{\label{fig3}(Color online) Order parameters of the approximate and exact system ground states for a 3-site resonator array, obtained from numerical diagonalization of the effective Hamiltonian $H^{eff}$ (black solid line) and the full Hamiltonian $H$ in Eq.\,(1) (red crosses), respectively. The parameters we choose are $g=g_{c}$, $\Delta_{c}=10g_{c}$, and $\Delta\in[-5g, 10g]$. }
 \end{figure}

 The increase in the number of sites will lead to a sharper phase transition \cite{reve12}. Suffering from the exponentially growing Hilbert space with the resonator array size, the exact matrix diagonalization method we use here can only obtain the solutions for very small resonator arrays. To give some primitive hints for the relation between the sharpness of the phase transition and the resonator array size, we compare the order parameters of the system ground states for 2-site, 3-site and 4-site resonator arrays in Fig.\,4. The parameters we choose are $g=g_{c}$, $\Delta_{c}=10g_{c}$, and $\Delta\in[-5g, 10g]$. For simplicity, the Hamiltonian we use is the effective Hamiltonian $H^{eff}$. It is shown that, the MI-SF transition takes place at the almost same value of $\Delta$ for different resonator arrays, and we may observe a more abrupt phase transition as the number of sites is increased.
 
 \begin{figure}[!h]
 \includegraphics[scale=0.63]{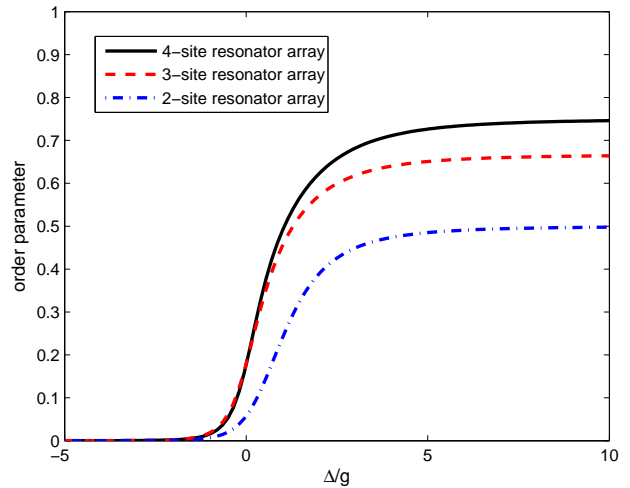}
 \setlength{\belowcaptionskip}{-0.4cm}
 \caption{\label{fig4}(Color online) Order parameters of the system ground states for 2-site (blue dashdotted line), 3-site (red dashed line) and 4-site (black solid line) resonator arrays. The parameters we choose are $g=g_{c}$, $\Delta_{c}=10g_{c}$, and $\Delta\in[-5g, 10g]$. }
 \end{figure}
 
 As presented avove, with a small size system of our proposed architecture, we may observe the main characters of MI-SF transition experimentally. In next section, we will discuss the related experimental issues.

 \section{dynamical observation of the photonic MI-SF transition}

 In the following, we give a concrete experimental procedure to engineer photonic MI-SF transition dynamically using our proposed architecture. We will describe in detail
 the total manipulation process (initialization, evolution and measurement) and analyze the various experimental imperfections.

 A simple scheme to initialize the system is the resonant pumping approach \cite{reve8,reve20}. We start in the MI regime with the total system in its absolute ground
 state $\otimes_{i=1}^{n}|0,g\rangle_{i}$. Then, by applying a global external microwave $\pi$-pulse, we are able to drive the filling factor of each resonator from zero
 to one and prepare the system in the required state $\otimes_{i=1}^{n}|1,-\rangle_{i}$. With the current circuit QED experimental techniques, this preparation method has
 the high fidelity larger than 99\%. 
 
 Now, we illustrate how to measure the variance of the polariton number in the resonator $R_{i}$. Our procedure utilizes two harmonic modes of $R_{i}$ which are engineered to have very different quality factors \cite{reve23}. As shown in Fig.\,1(b), $R_{i}$'s full-wave mode with resonance frequency $2w$ has an electric field antinode at the resonator center, and is hence strongly coupled to the measurement line. By choosing big coupling capacitances for the measurement line, the full-wave mode has a low quality factor. Conversely, $R_{i}$'s half-wave mode with resonance frequency $w$ has an electric field node at the resonator center and couple weakly to the measurement line, leaving the quality factor of this mode limited only by internal losses. To obtain $var(N_{i})$, we only need to get the probability distribution $p_{_{l}} \; (l=0,1,2,\,\ldots\, ,n)$ of the polariton number, with $p_{_{l}}$ the probability of finding exactly $l$ polaritons in $R_{i}$. In the experiment, one can obtain the value of $p_{_{l}}$ as follows: (1) First, switch off the effective polaritonic hopping instantaneously to isolate the system state from further evolution (this can be achieved by tuning the detuning $\Delta_{c}$ to a very large value instantaneously). (2) Adjust the detuning $\Delta$ adiabatically such that $\Delta/g\simeq5$. At this detuning, the polaritons are transferred into microwave photons, i.\,e.\,$|n,-\rangle_{i}\approx-|n,g\rangle_{i}$, and the qubit frequency is strongly shifted depending on the number of photons in the resonator \cite{reve21}. (3) Then, we drive transmon $i$ at the frequency $(\epsilon+2l\frac{g^{2}}{\Delta})$, thus to selectively populate the qubit into the excited state $|e\rangle_{i}$ if there are $l$ photons in the resonator. (4) Next, tune the qubit transition frequency $\epsilon$ such that transmon $i$ is decoupled from $R_{i}$'s half-wave mode but dispersively coupled to $R_{i}$'s full-wave mode. Then, the state of transmon $i$ can be measured fast by applying a microwave field $U_{i}^{m}$ of frequency $2w$ at the input port of the measurement line \cite{reve23}. (5) Repeat steps (1)-(4) for a large number of times $M$, and obtain the number of times $M_{l}$ in which transmon $i$ is excited into $|e\rangle_{i}$. Provided that we have always prepared the same system state before every implementation of steps (1)-(4), $p_{_{l}}$ can be obtained as $p_{_{l}}\approx\frac{M_{l}}{M}$. Finally, we can get the order parameter as
 \begin{equation}
 var(N_{i})=\sum_{l=0}^{n}l^{2}p_{_{l}}-(\sum_{l=0}^{n}lp_{_{l}})^{2}.
 \end{equation}
 For our annular architecture, performing the transmission measurement of individual resonators requires fabricating bridges on the circuit. To avoid this experimental difficulty, we can improve the measurement procedure by measuring reflection amplitudes instead of transmission amplitudes \cite{reve22}.

 Using the proposed architecture, we are able to observe the dynamical quantum phase transition of microwave photons by adiabatically changing the photon hopping to the
 on-site repulsion ratio. Similar approach has been adopted in the optical lattice experiment \cite{reve3}. Now, we analyze the feasibility of this procedure by some rough calculations based on the practical experimental parameters. It has been shown that, to make the system always remain in the many-body ground state of the varying Hamiltonian, the timescale of tuning the system from the MI state to the SF state or vice versa should be comparable to the photon hopping time $\frac{1}{\kappa}$ \cite{reve8}. If we perform the experiment by fixing $\Delta_{c}$ at a certain value and varying $\Delta$, the photon hopping time $\frac{1}{\kappa}=\frac{\Delta_{c}}{g_{c}^{2}}$. Coupling strength $g_{c}=2\pi\times200$ $MHz$ has been realized experimentally in Ref.\,\cite{reve23}. For $\Delta_{c}\in[10g_{c},100g_{c}]$, the photon hopping time can be calculated as $\frac{1}{\kappa}=8\sim80$ $ns$. On the other hand, an estimate of the polariton lifetime $\tau_{p}=2$ $\mu s$ is reasonable for the practical situation \cite{reve23}. It is clear that the timescale of the many-body dynamics is much shorter than the decay time of the mixed photon-qubit excitations, so our proposal is feasible with present-day technology. 
 
 In the practical experimental realizations, some system disorders are unavoidable because of lithography errors. The difference of the qubit transition frequencies can be compensated by controlling the local magnetic fluxes $\Phi$. The variation of the resonator frequencies is relatively difficult to control. To solve this problem, we can compensate this variation by inserting a superconducting quantum interference device (SQUID) into the central conductor of each resonator to make its frequency tunable \cite{reve28}. However, such an injected SQUID will introduce anharmonicity into the resonator energy levels. As a result, the effective on-site repulsion is composed of
 two parts: one part comes from the local Jaynes-Cummings interaction, the other part comes from the introduced resonator anharmonicity. Denoting the on-site repulsion
 induced by the resonator anharmonicity with $U^{'}_{eff}$, the critical point for the phase transition will be changed from $\frac{\kappa}{U_{eff}(1)}\simeq0.28$ to 
 $\frac{\kappa}{U_{eff}(1)+U^{'}_{eff}}\simeq0.28$. To observe the MI-SF transition, $U^{'}_{eff}$ has to fulfill the condition $U^{'}_{eff}<\frac{\kappa}{0.28}$. With the current fabrication techniques, the coupling strengths typically have larger relative fluctuations than the resonator frequencies. The variation in $g$ is harmless because the on-site repulsion $U_{eff}(1)$ is a complicated function of $g$ and $\Delta$, and the errors in $U_{eff}(1)$ induced by the deviation of $g$ can be compensated by individually adjusting the detuning $\Delta$. However, if the coupling strength $g_{c}$ fluctuates from sample to sample, the ac Stark shifts of the resonator frequencies in Eq.\,(3) can not be removed by going to the interaction picture. These terms are equivalent to additional parameter spreads in the resonator frequencies. Consequently, both the deviations in the resonator frequencies and the coupling strengths can be compensated by making the resonators tunable except it introduces additional part of effective on-site repulsion $U^{'}_{eff}$.
 
 In conclusion, we propose to observe the dynamical quantum phase transition of light using a 1-D array of coupled superconducting TLRs. The unique feature of our architecture is the good tunability of the photon hopping rate and the effective on-site repulsion. The local statistical property of each superconducting resonator can be analyzed readily using existing microwave techniques. With the recent progress in the multi-resonator experiments \cite{reve29}, our proposal may serve as a guide to coming experiments of quantum phase transition based on a small-scale resonator array.

 This work was supported by the Foundation for the Author of National Excellent Doctoral Dissertation of China (Grant No.\,200524), the Program for New Century Excellent Talents of China (Grant No.\,06-0920), and the National Natural Science Foundation of China (Grant No.\,11074307).

\end{document}